\documentclass[%
 preprint,
 amsmath,amssymb,
 aps,
]{revtex4-1}
\usepackage{natbib}
\usepackage{graphicx}
\usepackage{dcolumn}
\usepackage{bm}

\usepackage{subcaption}
\usepackage[dvipsnames]{xcolor}

\usepackage{caption}
\usepackage[labelformat = empty,position=top]{subcaption}
\usepackage[export]{adjustbox}
\usepackage{picture}

\begin{document}
\title{Mapping Phonon Modes from \\ Reduced-Dimensional to Bulk Systems}
\author{Hyun-Young Kim}
\author{Kevin D. Parrish}
\author{Alan J. H. McGaughey} \email{mcgaughey@cmu.edu} 

\affiliation{Department of Mechanical Engineering, Carnegie Mellon University, Pittsburgh, PA 15213 USA}

\date{\today}

\begin{abstract}

An algorithm for mapping the true phonon modes of a film, which are defined by a two-dimensional (2D) Brillouin zone, to the modes of the corresponding bulk material, which are defined by a three-dimensional (3D) Brillouin zone, is proposed.
The algorithm is based on normal mode decomposition and is inspired by the observation that the atomic motions generated by the 2D eigenvectors lead to standing-wave-like behaviors in the cross-plane direction.
It is applied to films between two and ten unit cells thick built from Lennard-Jones (LJ) argon, whose bulk is isotropic, and graphene, whose bulk (graphite) is anisotropic.
For LJ argon, the density of states deviates from that of the bulk as the film gets thinner due to phonon frequencies that shift to lower values.
This shift is a result of transverse branch splitting due to the film's anisotropy and the emergence of a quadratic acoustic branch.
As such, while the mapping algorithm works well for the thicker LJ argon films, it does not perform as well for the thinner films as there is a weaker correspondence between the 2D and 3D modes.
For graphene, the density of states of even the thinnest films closely matches that of graphite due to the inherent anisotropy, except for a small shift at low frequency.
As a result, the mapping algorithm works well for all thicknesses of the graphene films, indicating a strong correspondence between the 2D and 3D modes.

\end{abstract}
\maketitle
\section{\label{sec-intro}Introduction}

Phonons are quanta of energy associated with the atomic vibrations in a crystalline solid.
They provide a fundamental framework that is used for predicting and understanding a variety of phenomena, including thermal energy storage, thermal expansion, and thermal transport \cite{chen_nanoscale_2005, dove_introduction_2010}.
Every phonon mode is defined by a wave vector, which specifies its location in reciprocal space [i.e., the Brillouin zone (BZ)], polarization vector, and frequency.
The phonon modes represent a Fourier transform of the atomic positions to a coordinate system composed of independent harmonic oscillators \cite{ziman_electrons_2001}.

Advances in fabrication technologies have enabled the nanostructuring of three-dimensional (3D) bulk materials into lower-dimensional objects such as films [two-dimensional (2D)], wires [one-dimensional (1D)], and dots [zero-dimensional (0D)] \cite{kaiser_review_2002, xia_one-dimensional_2003, pu_colloidal_2018}.
Changes in the material dimensionality also affect the dimensionality of the BZ, thereby limiting the allowed wave vectors. 
This reduction in BZ dimensionality has been exploited to manipulate the properties of electrons in applications such as field-effect transistors, solar cells, and lasers \cite{bimberg_quantum_2011,mohammad_understanding_2014}.

Our objective is to probe how the dimensionality change impacts phonons in films, which are periodic in their two in-plane directions. 
For example, it is known that as a film's thickness is decreased, phonons begin to scatter with its boundaries in addition to intrinsic scattering mechanisms (i.e., with other phonons), which reduces their mean free paths \cite{liu_phononboundary_2004, turney_-plane_2010, jain_phonon_2013,cuffe_reconstructing_2015}. Films also limit the allowed cross-plane wave vectors \cite{turney_-plane_2010,wang_computational_2015} and generate surface phonons \cite{de_wette_study_1991}.
In contrast, flexural modes in single-layer graphene have a dramatically reduced scattering phase space compared to multi-layer graphene due to the emergence of new symmetries that limit their interactions \cite{lindsay_flexural_2010, balandin_superior_2008}.

\begin{figure}
    \begin{picture}(.49\textwidth,100)
	    \put(0,0){\includegraphics[width=.49\textwidth]{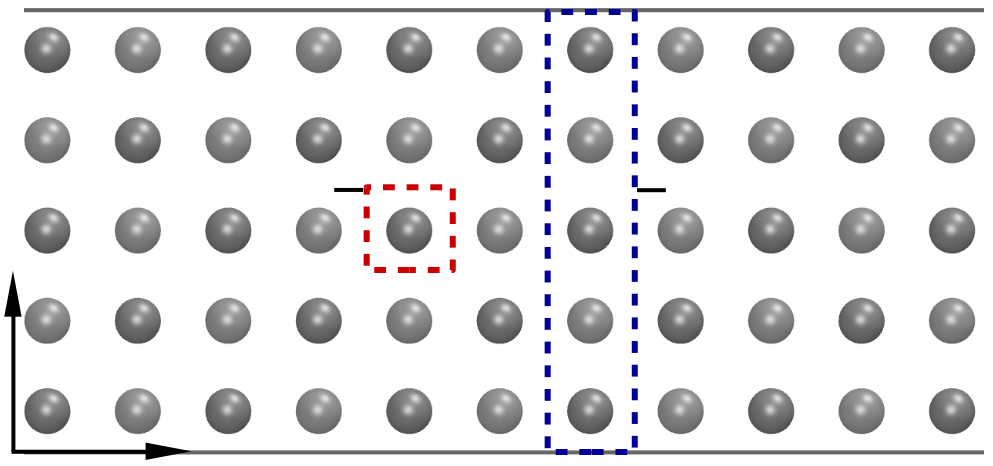}}
	    \put(.03\textwidth,.131\textwidth){{\fontsize{10pt}{11pt}\selectfont
3D Treatment}}
	    \put(.33\textwidth,.131\textwidth){{\fontsize{10pt}{11pt}\selectfont2D Treatment}}
	    \put(.01\textwidth,.087\textwidth){{\fontsize{10pt}{11pt}\selectfont \textit{z} (Cross-plane)}}
	    \put(.100\textwidth,-.012\textwidth){{\fontsize{10pt}{11pt}\selectfont \textit{x} (In-plane)}}
	\end{picture}
    \caption{Model thin film. The horizontal lines denote the film boundaries. The 3D treatment (red) uses the bulk unit cell, which allows cross-plane wave vectors. The 2D treatment (blue) has a unit cell that spans the film thickness, which disallows cross-plane wave vectors but creates additional polarizations.}
    \label{fig-treatment}
\end{figure}
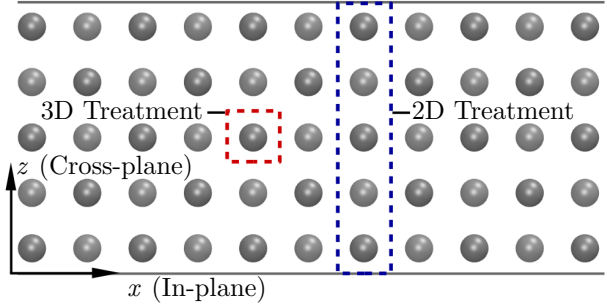

The phonon properties of a film can be predicted using the slab method, which uses a 2D BZ with a unit cell that spans the film thickness, as shown in Fig.~\ref{fig-treatment} \cite{de_wette_study_1991}.
The slab method is commonly used to model surface phonons \cite{de_wette_study_1991} and to predict the phonon properties of 2D materials such as multi-layer graphene \cite{lindsay_flexural_2011}.
The slab method stands in contrast to a common practice of predicting film phonon properties by using the modes corresponding to the 3D bulk unit cell (see Fig.~\ref{fig-treatment}) and assumptions related to the allowed cross-plane wave vectors and/or boundary scattering  \cite{turney_-plane_2010,jain_phonon_2013}.

While the 2D slab method is rigorous, modeling phonon scattering in this framework is computationally challenging. 
For example, if three-phonon scattering events are to be included, the required calculations scale with the fourth power of the number of atoms in the unit cell.
Furthermore, any required {\em ab initio} calculations with density functional theory will scale with the third power of the number of atoms in the computational cell.
This challenge leads to the question of when an approximate 3D representation is sufficient to describe the phonon modes in a film. 

We will investigate this question by developing and applying a mapping algorithm that links the 2D and 3D modes. 
The majority of previous work on mapping phonon modes, which is also called "unfolding," has focused on the link between an alloy and its virtual crystal representation \cite{boykin_approximate_2007,boykin_brillouin_2014,ikeda_mode_2017}, both of which are 3D systems.
Allen et al.~developed a general unfolding algorithm that they applied to (i) phonons in a diatomic 1D chain described by an exact two-atom basis and an approximate one-atom basis (i.e., the corresponding virtual crystal) and (ii) electrons in structures that are finite in one dimension (i.e., slabs) \cite{allen_recovering_2013}. 
Their slab calculation, however, requires the assumption of an artificial periodicity in the cross-plane direction to accommodate their unfolding formulation, which is periodic in nature.
Our proposed mapping algorithm eliminates the cross-plane periodicity requirement by searching for signatures of 3D phonon modes in the 2D phonon mode eigenvectors using normal mode decomposition \cite{ladd_lattice_1986,mcgaughey_quantitative_2004}.

The proposed algorithm is applied to films of Lennard-Jones (LJ) argon (whose bulk form is isotropic) and graphene (whose bulk form, graphite, is anisotropic).
The material details, underlying harmonic lattice dynamics theory, and mapping algorithm are described in Sec.~\ref{sec-method}.
The 2D and 3D density of states (DOS) and heat capacities are compared in Sec.~\ref{sec-dos}.
In Sec.~\ref{sec-disp}, the 2D film dispersions are mapped onto the 3D bulk dispersions to assess the algorithm and to identify similarities and differences between the 2D and 3D modes.
The differences for the LJ argon system reveal a transition from an isotropic bulk material to an anisotropic film.
This transition results in the separation of pairs of degenerate transverse branches into a cross-plane flexural branch and an in-plane transverse branch. This separation is naturally present in graphene, which is inherently anisotropic.

\section{\label{sec-method}Methodology}

\subsection{\label{sec-materials} Materials}

We model argon using the 12-6 LJ potential
\begin{equation} \label{eq-LJ}
\phi(r_{ij}) = 4\epsilon_{\mathrm{LJ}}\left[\left(\frac{\sigma_{\mathrm{LJ}}}{r_{ij}}\right)^{12}-\left(\frac{\sigma_{\mathrm{LJ}}}{r_{ij}}\right)^6\right],
\end{equation}
where $\phi(r_{ij})$ is the potential energy between atoms $i$ and $j$, $r_{ij}$ is the distance between them, and $\sigma_\mathrm{LJ}$ and $\epsilon_\mathrm{LJ}$ are the LJ length and energy scales ($\sigma_\mathrm{LJ} = 3.4 \times 10^{-10} \, \mathrm{m}$ and $\epsilon_\mathrm{LJ} = 1.67 \times 10^{-21} \, \mathrm{J} $ \cite{ashcroft_solid_1976}).
A cutoff of $2.5\sigma_\mathrm{LJ}$ is applied to limit the interaction range.
Bulk argon is a face-centered cubic crystal with a one-atom primitive unit cell and a zero-temperature, zero-pressure lattice constant of $5.269 \, \mathrm{\AA} $ \cite{mcgaughey_phonon_2004}. 
We use the four-atom conventional unit cell due to its simple-cubic structure, which has three orthogonal lattice vectors.
This choice results in the cross-plane direction of the films aligning with one of the lattice vectors and the other two lattice vectors lying in the same plane as the in-plane directions. The alignment of the lattice vectors and film directions allows for straightforward visualizations and matches the setup of the graphene system.
The in-plane directions are labeled with the Cartesian coordinates $x$ and $y$, and the cross-plane direction is labeled as $z$.
The lattice vectors, basis vectors, and unit cell are shown in Table \ref{T-lattice}.
In subsequent sections, the LJ argon films are identified by their thickness in bulk conventional unit cells.
An $N$-unit cell film therefore contains $4N$ atoms in its unit cell.

The atomic interactions within a graphene layer are modeled using the optimized Tersoff potential developed by Lindsay and Broido \cite{lindsay_optimized_2010}.
The in-plane carbon atoms take on a honeycomb structure with lattice constant of $2.50 \, \mathrm{\AA}$ \cite{lindsay_optimized_2010}.
The interlayer coupling in multi-layer graphene and graphite is modeled by a 12-6 LJ potential with $\sigma_\mathrm{LJ} = 3.276 \times 10^{-10} \, \mathrm{m}$ and $\epsilon_\mathrm{LJ} = 7.37 \times 10^{-22} \, \mathrm{J}$ \cite{lindsay_flexural_2011}.
The layer separation is $3.35 \, \mathrm{\AA}$ \cite{lindsay_flexural_2011} and a cutoff of $6.5 \, \mathrm{\AA}$ was applied to ensure that a layer only interacts with its neighboring layers. 
The layers are stacked in an AB structure that repeats every two layers to form a four-atom unit cell for graphite.
The structure is shown in Table \ref{T-lattice}.
The thicknesses of the multi-layer graphene films are defined by the number of bulk unit cells.
An $N$-unit cell graphene film therefore contains $4N$ atoms in its unit cell.
This choice of unit cell allows the same labeling scheme for the in-plane and cross-plane directions as for LJ argon. For both the LJ argon and multi-layer graphene films, a vacuum region with a size greater than the interaction cutoff is placed above and below the top and bottom surfaces.

The mapping algorithm (Sec.~\ref{sec-mapping}) was designed with the assumption of a uniform cross-plane layer separation.
Relaxing a film results in a non-uniform layer separation due to the non-symmetrical forces near its surfaces \cite{Bo2018Confinement}. We determined the variation in layer separation by relaxing films using energy minimization. For the five unit cell LJ argon film, the separation of the two outermost layers is 2\% larger than that at the film center. For the five unit cell graphene film, the separation of the two outermost layers is 1\% smaller than that at the film center. We believe that these variations are small enough that the impact on the phonon modes will not be significant. As such, the LJ argon and graphene films were not relaxed.

\begin{table*}
\caption{\label{T-lattice} Crystal structures for LJ argon (conventional unit cell) and graphite.}
\begin{tabular}{c|c|c}

\hline
\hline
&LJ Argon & Graphene and Graphite \\
\hline
Lattice vectors & $\mathbf{v}_1 = [a,0,0]$, $\mathbf{v}_2 = [0,a,0]$,&$\mathbf{v}_1 =[a,0,0]$, $\mathbf{v}_2 = [\frac{a}{2},\frac{a\sqrt{3}}{2},0]$,\\
& $\mathbf{v}_3 = [0,0,a]$& $\mathbf{v}_3=[0,0,c]$\\
\hline
Basis vectors & $\mathbf{b}_1=[0,0,0]$, $\mathbf{b}_2=[0,\frac{a}{2},\frac{a}{2}]$   &$\mathbf{b}_1=[0,0,0]$, $\mathbf{b}_2=[\frac{a}{2},\frac{a\sqrt{3}}{6},0]$\\
&$\mathbf{b}_3=[\frac{a}{2},\frac{a}{2},0]$, $\mathbf{b}_4=[\frac{a}{2},0,\frac{a}{2}]$&$\mathbf{b}_3=[0,\frac{a\sqrt{3}}{3},\frac{c}{2}]$, $\mathbf{b}_4=[\frac{a}{2},\frac{a\sqrt{3}}{6},\frac{c}{2}]$  \\
\hline
Lattice constant(s)&$a=5.269 \, \mathrm{\AA}$ \cite{mcgaughey_phonon_2006-1}&$a=2.50 \, \mathrm{\AA}$ \cite{lindsay_optimized_2010}, $c=6.70 \, \mathrm{\AA}$ \cite{lindsay_flexural_2011} \\
\hline
Unit cell && \\
 &
 \begin{picture}(100,130)
    \put(0,30){\includegraphics[width=110pt]{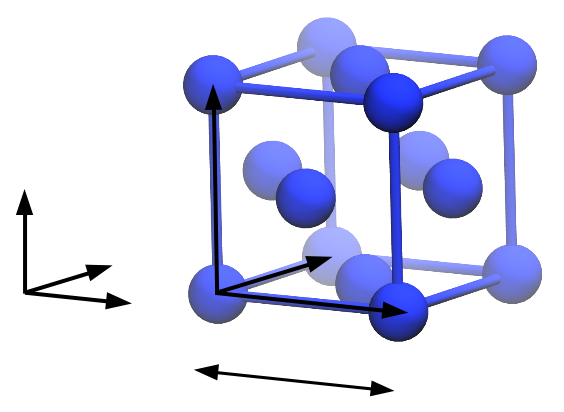}}
    \put(53,42){$\mathbf{v}_1$}
    \put(44,60){$\mathbf{v}_2$}
    \put(30,70){$\mathbf{v}_3$}
    \put(52,30){$a$}
    \put(15,42){$x$}
    \put(10,60){$y$}
    \put(-5,70){$z$}

 \end{picture}
 & 
 \begin{picture}(100,130)
    \put(0,0){\includegraphics[width=110pt]{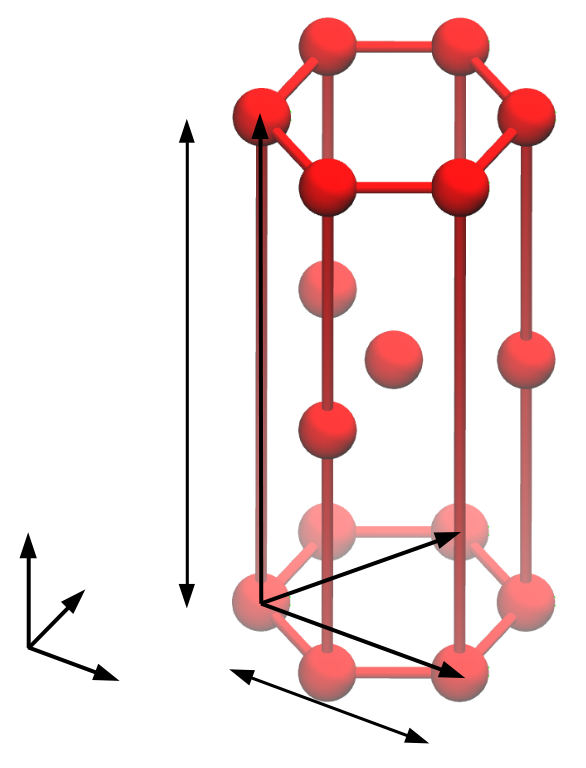}}
    \put(73,25){$\mathbf{v}_1$}
    \put(73,33){$\mathbf{v}_2$}
    \put(36,73){$\mathbf{v}_3$}
    \put(55,5){$a$}
    \put(26,70){$c$}
    \put(15,10){$x$}
    \put(10,37){$y$}
    \put(-4,41){$z$}

\end{picture}
 \\
 \hline
 $\kappa_z=0$ BZ Slice&& \\
 &
 \begin{picture}(100,130)
    \put(0,30){\includegraphics[width=110pt]{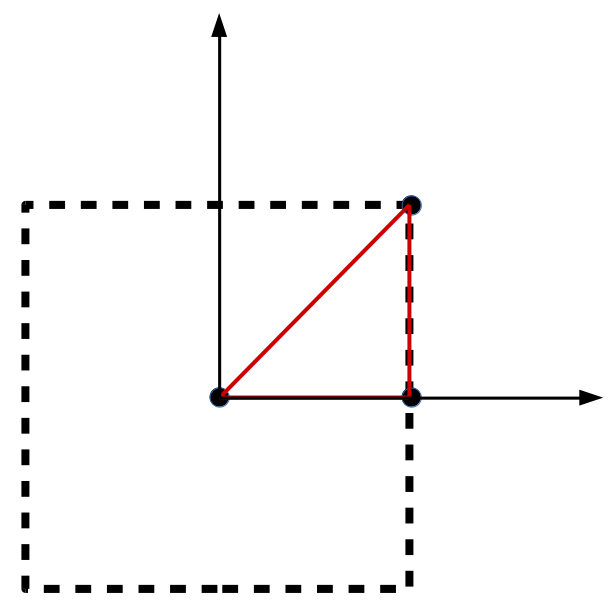}}
    \put(95,60){$\kappa_x$}
    \put(27,125){$\kappa_y$}
    \put(30,58){$\mathrm{\mathbf{\Gamma}}$}
    \put(77,57){$\mathrm{\mathbf{X}}$}
    \put(77,100){$\mathrm{\mathbf{M}}$}
    \put(-10,16){$\mathrm{\mathbf{\Gamma}} = [0,0,0], \, \mathrm{\mathbf{X}} = [\frac{\pi}{a},0,0],$}
    \put(-10,0){$\mathrm{\mathbf{M}} = [\frac{\pi}{a},\frac{\pi}{a},0]$}

 \end{picture}
 & 
 \begin{picture}(100,130)
    \put(0,27){\includegraphics[width=110pt]{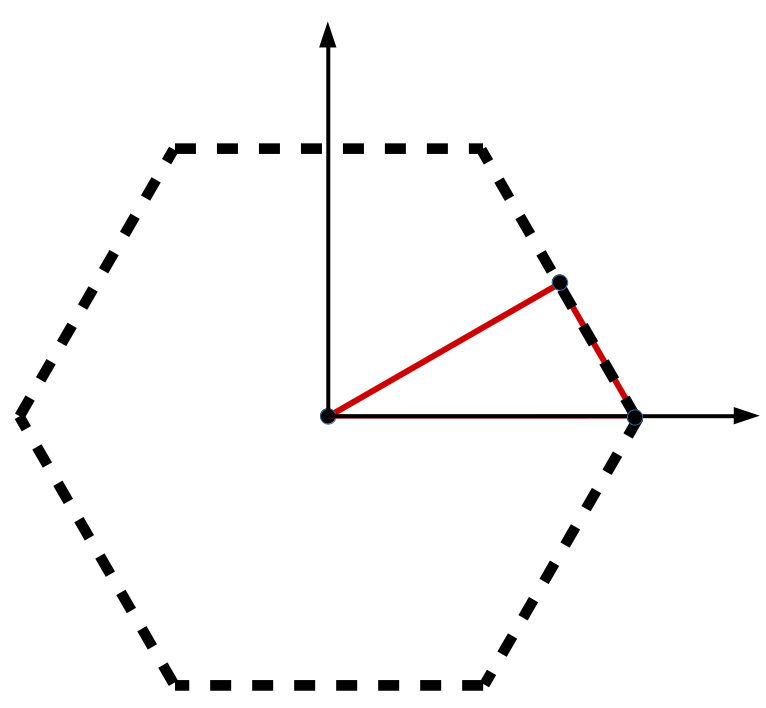}}
    \put(105,60){$\kappa_x$}
    \put(34,120){$\kappa_y$}
    \put(42,58){$\mathrm{\mathbf{\Gamma}}$}
    \put(92,56){$\mathrm{\mathbf{K}}$}
    \put(84,90){$\mathrm{\mathbf{M}}$}
    \put(-10,16){$\mathrm{\mathbf{\Gamma}} = [0,0,0], \, \mathrm{\mathbf{K}} = [\frac{4\pi}{3a},0,0],$}
    \put(-10,0){$\mathrm{\mathbf{M}} = [\frac{\pi}{a},\frac{\pi\sqrt{3}}{3a},0]$}

\end{picture}
 \\

 \hline
 \hline
\end{tabular}
\end{table*}

\subsection{\label{sec-hld} Harmonic Lattice Dynamics}
The frequencies $\omega$ and polarization vectors $\bf{e}$ of the phonon modes with the wave vector $\pmb{\kappa}$ can be found by solving the eigenvalue problem
\begin{equation} \label{eq-hld}
\omega^2(\pmb{\kappa},\nu)\bf{e}(\pmb{\kappa},\nu) = \bf{D}(\pmb{\kappa}) \bf{e}(\pmb{\kappa},\nu),
\end{equation}
where $\nu$ labels the polarization \cite{dove_introduction_2010}. $\bf{D}$ is the dynamical matrix, whose entries are
\begin{widetext}
\begin{equation}\label{eq-dynam}
\begin{split}
D_{3(j-1)+\alpha, 3(j'-1)+\beta}(jj',\pmb{\kappa}) &=\\
\frac{1}{(m_jm_{j'})^{1/2}} & \sum_{k'} \Phi_{\alpha\beta}(jk_0;j'k')\mathrm{exp}(i\pmb{\kappa}\cdot[\mathbf{r}(j'k')-\mathbf{r}(jk_0)]).
\end{split}
\end{equation}
\end{widetext}
Here, $m$ is the atomic mass and $\mathbf{r}$ is the equilibrium position of basis atom $j$ in unit cell $k$, with $k_0$ being the central unit cell.
The $\Phi_{\alpha\beta}$ terms are the second-order (i.e., harmonic) force constants, where $\alpha$ and $\beta$ denote the Cartesian directions [1, 2, and 3 for $x$, $y$, and $z$ in Eq.~\eqref{eq-dynam}].
They are defined as the second derivative of the system potential energy $U$ with respect to displacements $u$, or equivalently as the negative of the first derivative of the force $F^\alpha_{jk}$, as 
\begin{equation}\label{eq-fc}
    \Phi_{\alpha\beta}(jk;j'k') = \frac{\partial U}{\partial u_{jk}^\alpha \partial u_{j'k'}^\beta}= -\frac{\partial F^\alpha_{jk}}{\partial u_{j'k'}^\beta}.
\end{equation}
We obtain the harmonic force constants with a four-point central difference formula on the force.
Translational invariance is enforced using a Lagrangian approach \cite{li_thermal_2012}.
The harmonic lattice dynamics calculation can be performed either on a set of wave vectors evenly distributed in the BZ to obtain the DOS or along high-symmetry directions to obtain the dispersion.

\subsection{\label{sec-mapping}Mapping Algorithm}

Our mapping algorithm makes use of the normal mode decomposition technique, which maps atomic trajectories onto phonon (i.e., normal) modes \cite{ladd_lattice_1986,mcgaughey_quantitative_2004}.
While the atomic motions in a 2D film cannot form a traveling wave in the cross-plane direction due to a lack of periodicity, they may form standing waves.
These standing waves can then be interpreted as a superposition of two traveling waves with opposing cross-plane wave vectors that are associated with 3D phonons.

Each 2D phonon mode has a polarization vector that describes the displacements from equilibrium of the atoms in the unit cell.
We calculate the atomic motions of all atoms in the unit cell, $\mathbf{u}_{\mathrm{2D}}$, at a temporal phase $t$ as
\begin{equation}\label{eq-disp}
\mathbf{u}_{\mathrm{2D}}(\pmb{\kappa}_\mathrm{2D},\nu_\mathrm{2D},t) = \mathrm{Re}\{\mathbf{e}_{2D}(\pmb{\kappa}_\mathrm{2D},\nu_\mathrm{2D})\mathrm{exp}[i\omega_\mathrm{2D}(\pmb{\kappa}_\mathrm{2D},\nu_\mathrm{2D})t]\}.
\end{equation}
The temporal phase is varied in the range of $0 \leq t < \frac{2\pi}{\omega_{\mathrm{2D}}}$ in intervals of $\frac{\pi}{50\omega_{\mathrm{2D}}}$ to cover a full period.
Here, time is not a dynamic variable. Shifting the polarization vector by a temporal phase allows for the full range of atomic motions in that phonon mode to be explored, which helps to reduce numerical errors in the mapping algorithm.

Normal mode decomposition is then applied to project the atomic displacements at each $t$ onto all 3D wave vectors ($\pmb{\kappa}_\mathrm{3D}$) and their polarization vectors ($\mathbf{e}_\mathrm{3D}$). 
As we are working in orthogonal lattices and only mapping to determine the cross-plane wave vector component $\kappa_z$, the in-plane components $\kappa_x$ and $\kappa_y$ are assumed to be the same in the 2D and 3D BZs.
The amplitude of the 3D normal mode coordinate $q$ is calculated at each $t$ from 
\begin{widetext}
\begin{equation} \label{eq-map}
q(\pmb{\kappa}_\mathrm{3D},\nu_{3D},\nu_\mathrm{2D},t)  = \sum_{\alpha,j,l} \sqrt{\frac{m_j}{N}}u_{\mathrm{2D},\alpha,jl} ([\kappa_{\mathrm{3D},x},\kappa_{\mathrm{3D},y}],\nu_{2D},t)e_{\mathrm{3D},\alpha,jl}^{\dagger}(\pmb{\kappa}_\mathrm{3D},\nu_\mathrm{3D})\mathrm{exp}(-i\pmb{\kappa}_\mathrm{3D}\cdot [\mathbf{r}(jl)-\mathbf{r}(jl_0)]),
\end{equation}
\end{widetext}
where $u_{\mathrm{2D},\alpha,jl}$ is the $\alpha$-component of the displacement of basis atom $j$ in the 2D unit cell layer $l$.
The unit cell layer is defined as the index of the 3D unit cell stacked in the cross-plane direction in the 2D unit cell, with $l_0$ referring to the bottom of the 2D unit cell.
Similarly, $e_{\mathrm{3D},\alpha,jl}$ refers to the component of the polarization vector corresponding to movement of basis atom $j$ and unit cell layer $l$ in the $\alpha$-direction in the 3D description, with the $\dagger$ symbol indicating its complex conjugate.

The sampling of $\kappa_z$ is limited to
\begin{equation}\label{eq-destination}
\kappa_z = \frac{n\pi}{Na_{\mathrm{cross}}}, n = \pm\, 0,1,2, ..., N
\end{equation}
to only consider modes that can be described by the discrete number of bulk unit cells in the film thickness.
The cross-plane lattice constant $a_\mathrm{cross}$ is $a$ for argon and $c$ for graphene (Table \ref{T-lattice}).
The wavelength, $2\pi/\kappa_z$, can be at most twice the film thickness (i.e., at least half of the wavelength must fit into the film).
Eq.~\eqref{eq-destination} thus results in $N+1$ mapping destinations for the $N$ polarizations for a given $\pmb{\kappa}_\mathrm{2D}$.
As such, there will not be a complete one-to-one correspondence, as will be discussed in Sec. \ref{sec-disp}.

The normal mode coordinate magnitudes are then summed over the entire period for each 2D mode for every possible 3D cross-plane wave vector and polarization, generating quantities of the form
\begin{equation}\label{eq-qavg}
\overline{q}(\pmb{\kappa}_{\mathrm{3D}},\nu_{\mathrm{3D}},\nu_{\mathrm{2D}}) = \sum_t  \left | q(\pmb{\kappa}_\mathrm{3D},\nu_\mathrm{3D},\nu_\mathrm{2D},t) \right |.
\end{equation}
For each 2D mode, the summed normal-mode coordinate is plotted versus the cross-plane wave vector. 
The $\kappa_z$ with the maximum $\overline{q}$ is used as the mapping destination.

This application of normal mode decomposition to the 2D atomic trajectories is similar to the Bloch-wave unfolding algorithm described by Allen et al. \cite{allen_recovering_2013}.
Their unfolding approach, however, relies on comparing systems with equal dimensionality (i.e., 3D to 3D).
This reliance is due to their unfolding algorithm comparing a polarization vector to translated versions of itself, with periodicity necessary to wrap around the edges of a supercell.
Allen et al.~mitigated this limitation by introducing a midline to represent the surfaces of a film, with the polarization vectors reflecting across it with some corrections.
This strategy allowed them to unfold electrons in a silicon film to the primitive bulk unit cell.
For phonons, however, normal mode decomposition enables a direct comparison between the 2D and 3D polarization vectors, such that the midline approximation is unnecessary.

\section{\label{sec-results}Results}
\subsection{\label{sec-dos} Phonon Density of States}

\begin{figure*}[tb]
	\begin{tabular}{c c}
	\begin{subfigure}[t]{.03\textwidth}
    (a)
    \end{subfigure}
	\begin{subfigure}[t]{.42\textwidth}
	\includegraphics[width=\textwidth,valign=t]{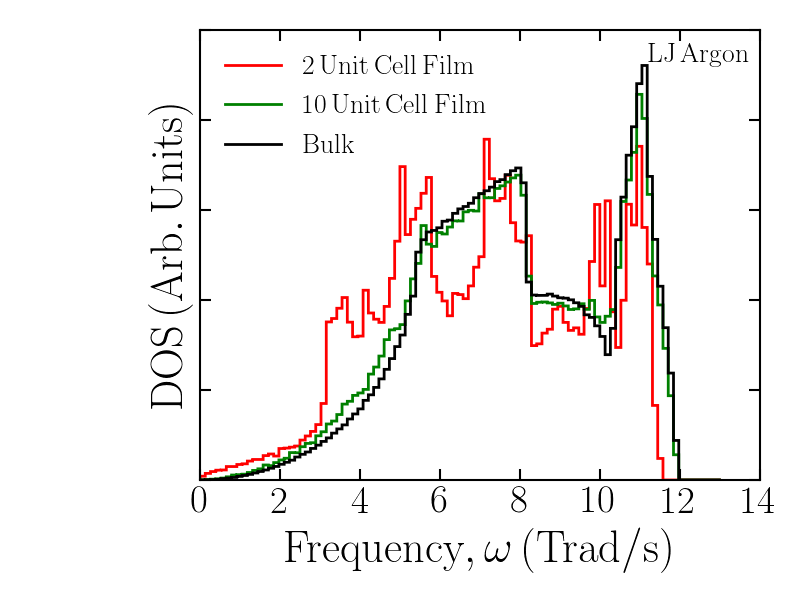}
    \end{subfigure}
    &
	\begin{subfigure}[t]{.03\textwidth}
    (b)
    \end{subfigure}
	\begin{subfigure}[t]{.42\textwidth}
	\includegraphics[width=\textwidth,valign=t]{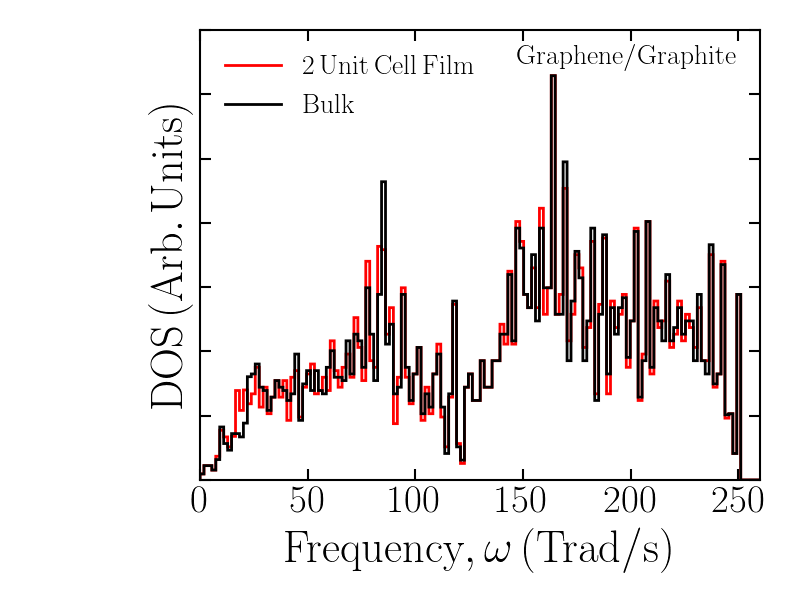}
    \end{subfigure}
    \\
	\begin{subfigure}[t]{.03\textwidth}
	(c)
    \end{subfigure}
	\begin{subfigure}[t]{.42\textwidth}
	\includegraphics[width=\textwidth,valign=t]{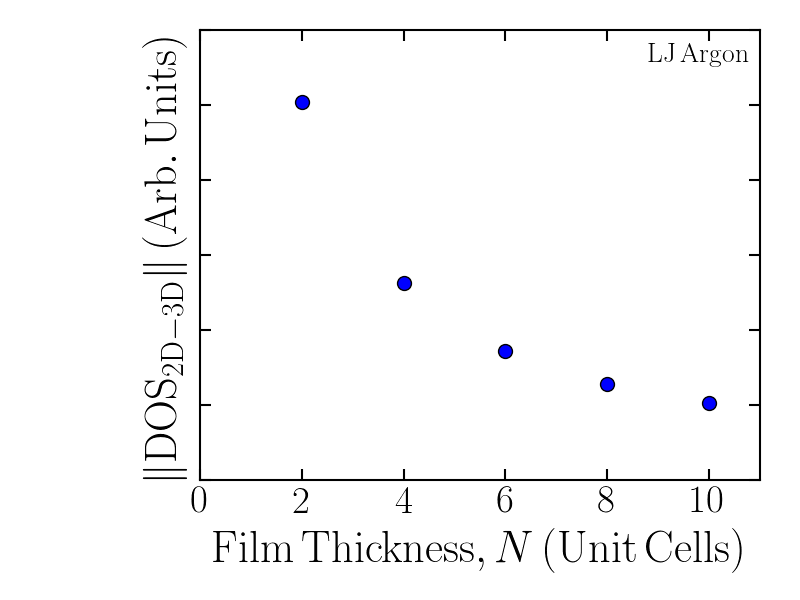}
    \end{subfigure}
    &
    \begin{subfigure}[t]{.03\textwidth}
    (d)
    \end{subfigure}
	\begin{subfigure}[t]{.42\textwidth}
	\includegraphics[width=\textwidth,valign=t]{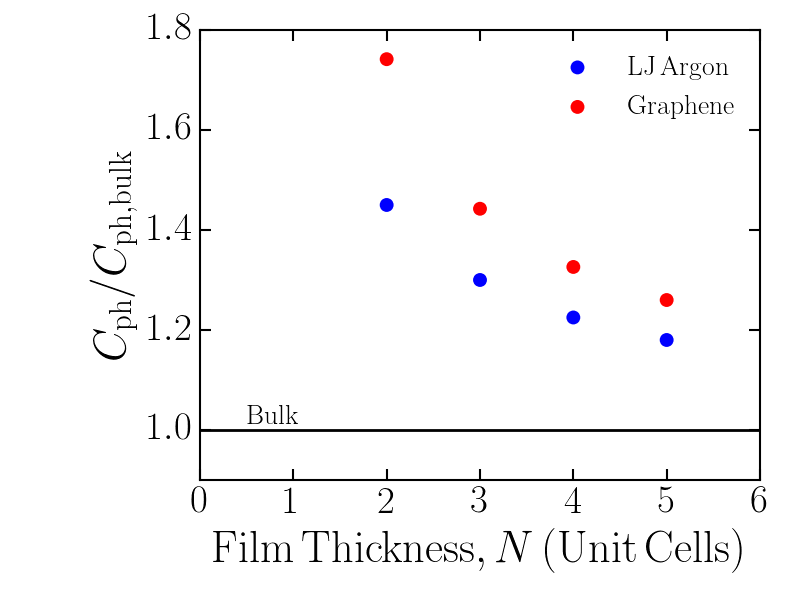}
    \end{subfigure}
    \end{tabular}
    \caption{
    (a) DOS of the two unit cell film, ten unit cell film, and bulk LJ argon systems. As a result of anisotropy, the thinner film DOS differs significantly from the bulk DOS.
    (b) DOS of the two unit cell graphene film and bulk graphite systems. Even at this small thickness, the two unit cell DOS matches the bulk DOS due to the weak cross-plane interactions.
    (c) Difference between the 2D and 3D DOS as a function of film thickness for LJ argon. As the thickness increases, the 2D DOS converges to the 3D DOS.
    (d) Specific heat as a function of film thickness, calculated at a temperature of 10 K. The 2D and 3D specific heats diverge as thickness decreases due to differences in the DOS at low frequencies.
    }
    \label{fig-dos}
\end{figure*}

Before applying the mapping algorithm, we first compare the 2D and 3D DOS for the LJ argon [Fig. \ref{fig-dos}(a)] and graphene [Fig. \ref{fig-dos}(b)] systems.
The in-plane wave vector sampling was the same for the 2D and 3D systems, at $100\times100$ for LJ argon and $44\times44$ for graphene.
The 3D BZ was further sampled with 100 cross-plane wave vectors for LJ argon and 44 cross-plane wave vectors for graphite.
The DOS were normalized to have the same integrated area to allow for a direct comparison between different thicknesses.

For LJ argon, the  two unit cell film DOS [Fig. \ref{fig-dos}(a)] differs significantly from the 3D DOS with a shift to lower frequencies. 
The ten unit cell film DOS, in comparison, is similar to the 3D DOS. 
We quantify the difference between the 2D and 3D DOS by calculating the Euclidean distance $\lVert \mathrm{DOS_{2D-3D}}\rVert$ between them as
\begin{equation}\label{eq-dosmetric}
    \lVert \mathrm{DOS_{2D-3D}} \Vert= \sqrt{\sum_i (h_{2D,i}-h_{3
    D,i})^2},
\end{equation}
where $i$ indicates the frequency bin and $h_i$ is the DOS for that bin.
A hundred uniform bins were used to divide the frequency range.
This quantity was calculated for two to ten unit cell LJ argon films and the results are plotted in Fig. \ref{fig-dos}(c). 
As the film thickness increases, the difference between the 2D and 3D DOS steadily decreases.
This result suggests that while the phonon modes of the two unit cell film may differ from those of the 3D system, at larger thicknesses the 2D and 3D modes may demonstrate a close to one-to-one correspondence.

In contrast, the 2D DOS of a graphene film is nearly identical to the bulk graphite DOS even at two unit cells thick, with small differences only at low frequencies.
The cross-plane bonding is much weaker than the in-plane bonding.
As such, the in-plane forces are not affected by the removal of layers and the majority of the frequencies are unchanged.
The acoustic flexural phonon modes are the most affected, which is reflected in the deviation of the low frequency region of the DOS where they reside. In contrast, the LJ argon films introduce anisotropy that is not present in the bulk system, which leads to differences between the 2D and 3D phonon modes.

The differences between the 2D and 3D DOS can be further examined by calculating the specific heat, $C_{\mathrm{ph}}$, from  \cite{togo_first-principles_2010}
\begin{equation}\label{eq-cph}
C_{\mathrm{ph}} = \frac{1}{V}\sum_{\pmb{\kappa},\nu} \frac{k_\mathrm{B}x^2e^x}{(e^x-1)^2}.
\end{equation}
Here, $x = \frac{\hbar\omega(\pmb{\kappa},\nu)}{k_\mathrm{B} T}$, $T$ is the temperature, $k_\mathrm{B}$ is the Boltzmann constant, $\hbar$ is the reduced Planck constant, and $V$ is the crystal volume.
The results for both materials at a temperature of 10 K are plotted as a function of film thickness in Fig. \ref{fig-dos}(d), where they are normalized by the corresponding bulk value.
At this low temperature, the specific heat is mainly determined by the low-frequency modes.
Since the films of both materials have DOS that deviate from bulk at low frequencies, their specific heats should also deviate.
This result is seen in both LJ argon (with a specific heat up to 1.45 times the bulk value) and graphene (with a specific up to 1.75 times the bulk value).

\subsection{\label{sec-disp} Phonon Dispersion}
\subsubsection{\label{sec-ljarg} LJ Argon}

\begin{figure*}
	\begin{tabular}{c c}
	\begin{subfigure}[t]{.03\textwidth}
	(a)
    \end{subfigure}
	\begin{subfigure}[t]{.42\textwidth}
	\includegraphics[width=\textwidth,valign=t]{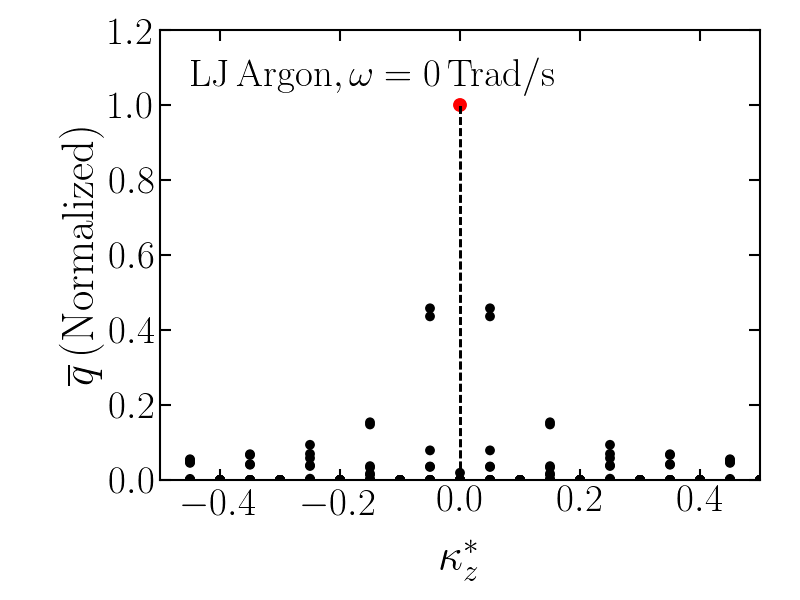}
    \end{subfigure}
    &
	\begin{subfigure}[t]{.03\textwidth}
	(b)
    \end{subfigure}
	\begin{subfigure}[t]{.42\textwidth}
	\includegraphics[width=\textwidth,valign=t]{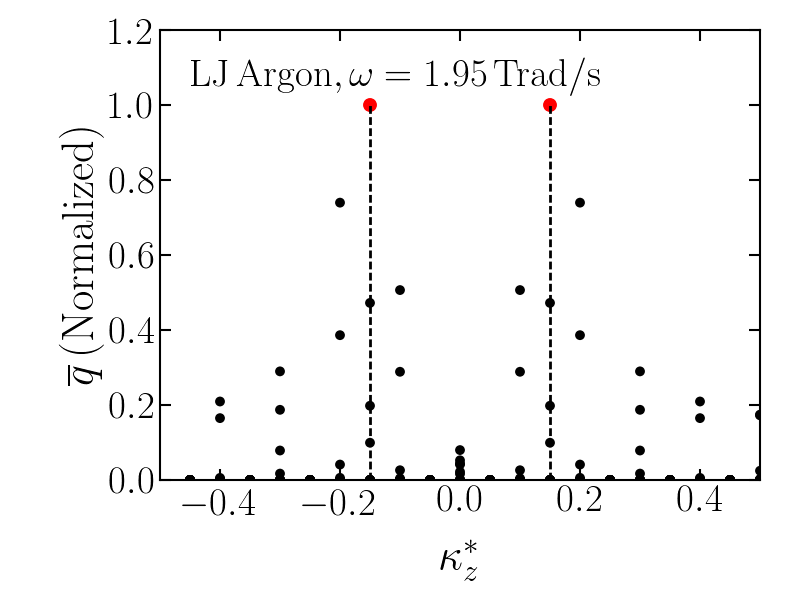}
    \end{subfigure}
    \\
	\begin{subfigure}[t]{.03\textwidth}
	(c)
    \end{subfigure}
	\begin{subfigure}[t]{.42\textwidth}
	\includegraphics[width=\textwidth,valign=t]{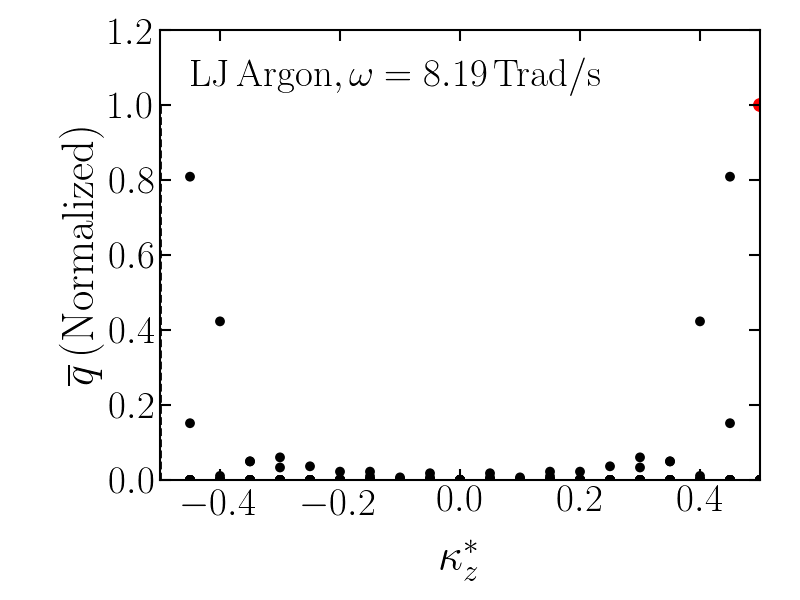}
    \end{subfigure}
    &
    \begin{subfigure}[t]{.03\textwidth}
    (d)
    \end{subfigure}
	\begin{subfigure}[t]{.42\textwidth}
	\includegraphics[width=\textwidth,valign=t]{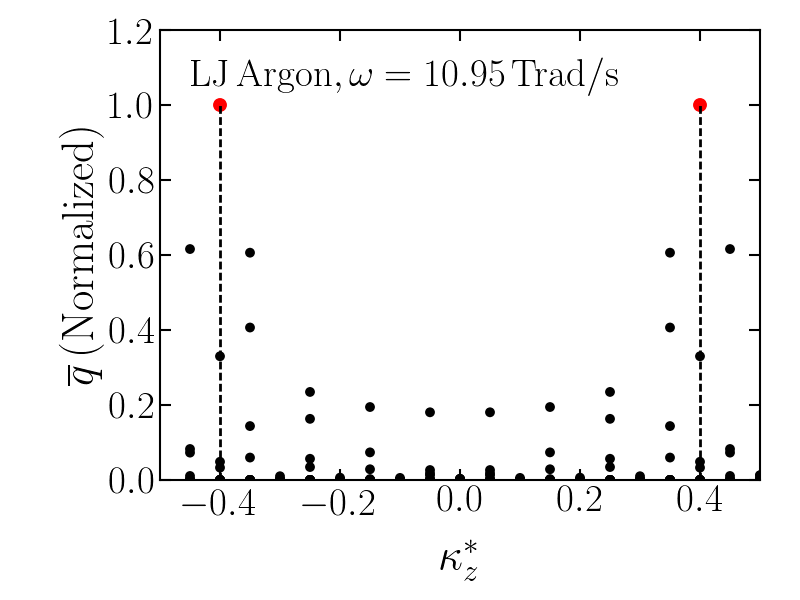}
    \end{subfigure}
    \end{tabular}
    \caption{$\overline{q}$ sweeps for four $\mathbf{\Gamma}$-point LJ argon modes in a ten unit cell film (i.e., $\kappa_x =\kappa_y = 0$). 
    The mapping algorithm is run for all 3D polarizations at each $\kappa_z$ value and the normalization is based on the maximum value.
    The peaks are marked red and are at (a) $|\kappa_z^*| = 0$, (b) $|\kappa_z^*| = 0.15$, (c) $|\kappa_z^*| = 0.50$, and (d) $|\kappa_z^*| = 0.40$.
    The peak locations are used to map between the 2D and 3D modes.}
    \label{ex-sweep}
\end{figure*}

The mapping algorithm was applied to LJ argon films with thicknesses of two to ten unit cells. 
Examples of the $\overline{q}$ sweep through $\kappa_z$ [Eq.~\eqref{eq-qavg}] are plotted in Figs.~\ref{ex-sweep}(a)-\ref{ex-sweep}(d) for a ten unit cell LJ argon film at the 2D $\mathrm{\mathbf{\Gamma}}$-point (i.e., the center of the BZ, where $\kappa_x=\kappa_y=0$) for four polarizations.
For simplicity, we define $\kappa_z^*=\frac{\kappa_z}{2\pi/a_{cross}}$.
The $\overline{q}$ sweeps are symmetrical about $\kappa_z^*=0$ for each frequency.
This result is expected, as propagating cross-plane modes cannot exist in a 2D film.
Any cross-plane periodicity must be captured by a standing wave built from two propagating waves that are traveling in opposite directions. 
There are clear maxima in $\overline{q}$ at each polarization, which are shown with dashed vertical lines and red markers.
These peaks identify the mapping location in the 3D BZ.

\begin{figure*}
     \includegraphics[width=\textwidth]{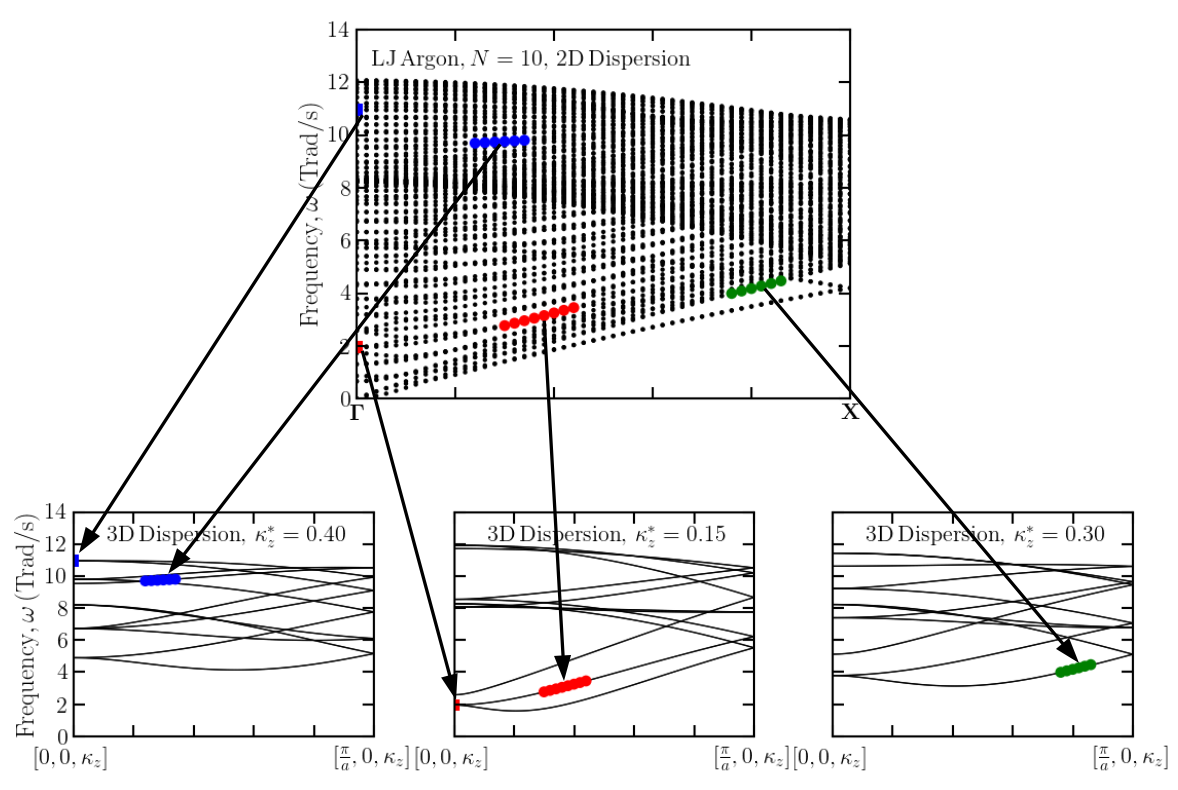}
    \caption{Mapping of selected LJ argon phonon modes from the 2D dispersion (top) to multiple 3D dispersions (bottom) in a 10 unit cell film.
    The modes from Figs. \ref{ex-sweep}(b) and \ref{ex-sweep}(d) are plotted as blue and red squares.
    }
    \label{fig-process}
\end{figure*}

To visualize the mapping algorithm, the 2D dispersion for the ten unit cell LJ argon film with varying $\kappa_x$ and $\kappa_y=0$ (i.e., along the $\mathbf{\Gamma}-\mathbf{X}$ direction) is plotted in Fig.!\ref{fig-process}. Plotted below the 2D dispersion are corresponding bulk 3D dispersions at $\kappa_z^*$ values of 0.15, 0.30, and 0.40.
Portions of the 2D dispersion are highlighted in red, green, and blue.
Applying the mapping algorithm matches these modes to $\kappa_z^*$ values of $0.15$ (red), $0.30$ (green), and $0.40$ (blue). 
The red and blue $\pmb{\Gamma}$-point modes correspond to the data from Figs.~\ref{ex-sweep}(b) and \ref{ex-sweep}(d).

\begin{figure*}	
	\begin{tabular}{c}
	\begin{subfigure}[t]{.03\textwidth}
	(a)
    \end{subfigure}
    
  	\begin{subfigure}[t]{.45\textwidth}
	\includegraphics[width=\textwidth,valign=t]{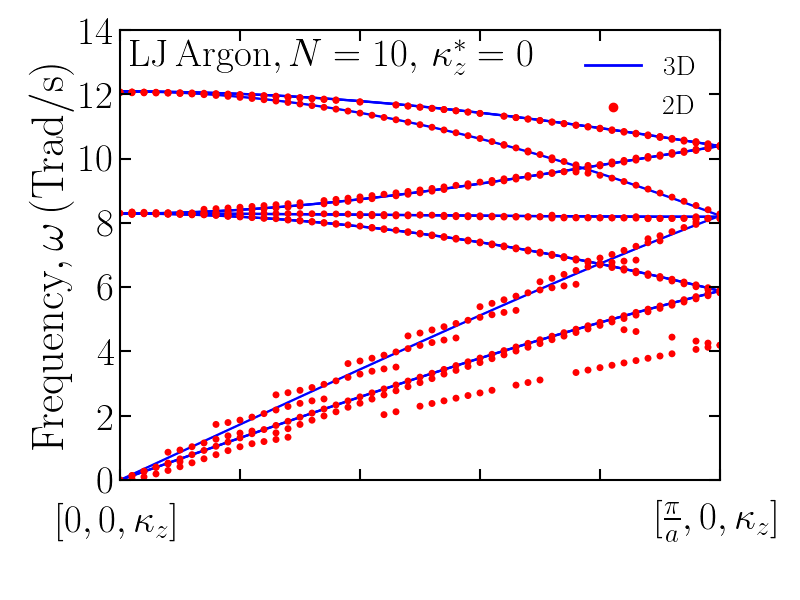}
    \end{subfigure}
   	\\
    
 	\begin{subfigure}[t]{.03\textwidth}
 	(b)
    \end{subfigure}
  	\begin{subfigure}[t]{.45\textwidth}
    \includegraphics[width=\textwidth,valign=t]{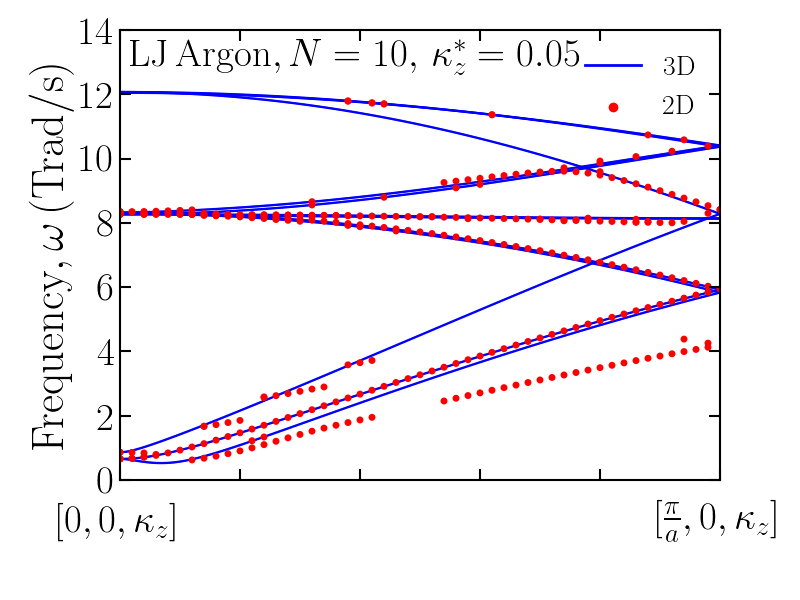}
    \end{subfigure}
    \\
    \begin{subfigure}[t]{.03\textwidth}
    (c)
    \end{subfigure}
    
  	\begin{subfigure}[t]{.45\textwidth}
	\includegraphics[width=\textwidth,valign=t]{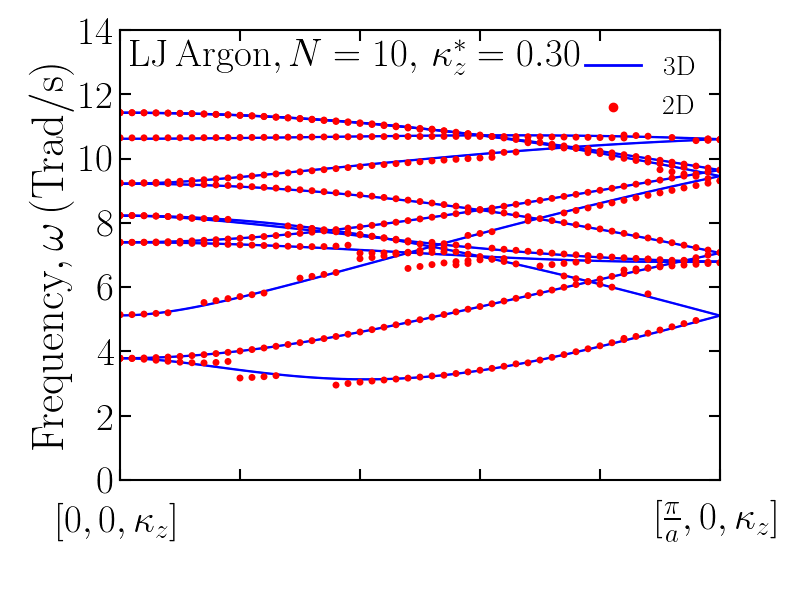}
    \end{subfigure}
    
    \end{tabular}
    
    \caption{LJ argon dispersions along $\kappa_x$ for a ten unit cell film with $\kappa_y=0$ and (a) $\kappa_z^*=0$, (b) $\kappa_z^*=0.05$, and (c) $\kappa_z^*=0.30$.}
    \label{argon-disp}
\end{figure*}

If there are strong similarities between the 2D and 3D phonon modes, then the mapped 2D dispersions should trace out the 3D branches.
The results for the ten unit cell LJ argon film are shown in Figs.~\ref{argon-disp}(a)-\ref{argon-disp}(c) from $[0,0,\kappa_z]$ to $[\pi/a,0,\kappa_z]$ at $\kappa_z^*$ values of 0, 0.05, and 0.30.
With the exception of $\kappa_z^*=0.05$, the mapped 2D modes mostly trace the 3D branches.
This result demonstrates that 2D modes can be mapped to 3D modes at specific cross-plane wave vectors.
There are some deviations in the acoustic branches for $\kappa_z^*=0$ and $\kappa_z^*=0.05$.
For example: (i) A 3D branch traced with small pieces of different 2D branches.
This piece-wise mapping happens because the shapes of the 2D branches differ from those of the 3D branches, such that multiple branches are required to map across the BZ.
(ii) A mapped 2D branch has frequencies lower than any 3D branch.

\clearpage

\begin{figure}
	\includegraphics[width=.45\textwidth,valign=t]{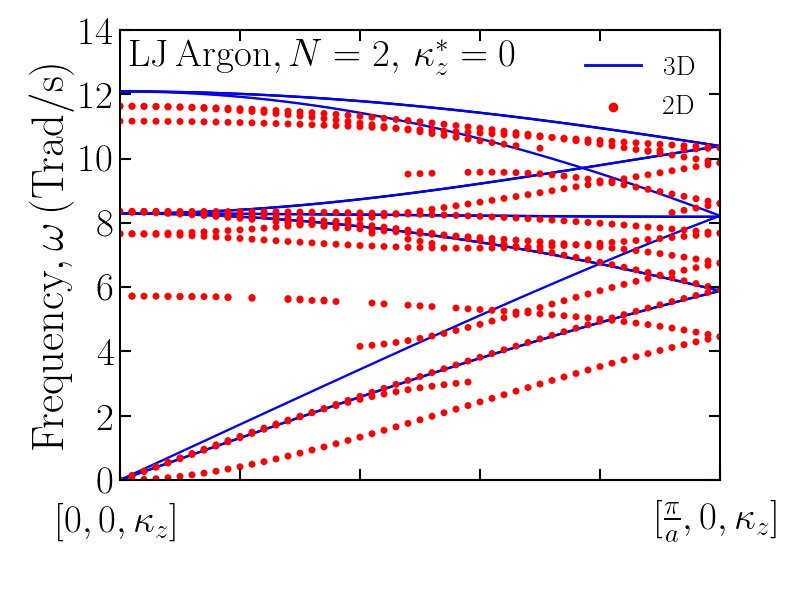}

    \caption{Dispersion of the two unit cell LJ argon film compared to the 3D dispersion at $\kappa_z^*=0$.
   The 2D dispersion has a downward frequency shift for all branches and a splitting of the degenerate bulk transverse branches due to its anisotropy.}
    \label{argon-lj1}
\end{figure}

The second discrepancy can be examined by considering the mapped dispersion along $\kappa_x$ for a two unit cell film for $\kappa_y=\kappa_z=0$, as shown in Fig. \ref{argon-lj1}.
The downward frequency shift is present in all branches and can be attributed to two phenomena.
First, there are additional branches (e.g., the optical branch with a $\pmb{\Gamma}$-point frequency near $6$ Trad/s).
Such branches are a result of the film anisotropy. 
As the film thickness decreases, the degenerate bulk transverse branches differentiate.
The atoms in one branch move in the in-plane direction, while the atoms in the other move in the cross-plane direction.
Due to the free surfaces at the top and bottom of the film, the restoring forces in the in-plane and cross-plane directions will be different, which leads to the branch splitting. 
This effect will occur for all degenerate acoustic and optical branches.
Second, one of the acoustic branches is quadratic, which is expected in a two-dimensional system.
A similar behavior will be present in the ten unit cell film and leads to a reduction in frequency.

\begin{figure}	
	\begin{tabular}{c}
	\begin{subfigure}[t]{.03\textwidth}
	(a)
    \end{subfigure}
    
  	\begin{subfigure}[t]{.45\textwidth}
	\includegraphics[width=\textwidth,valign=t]{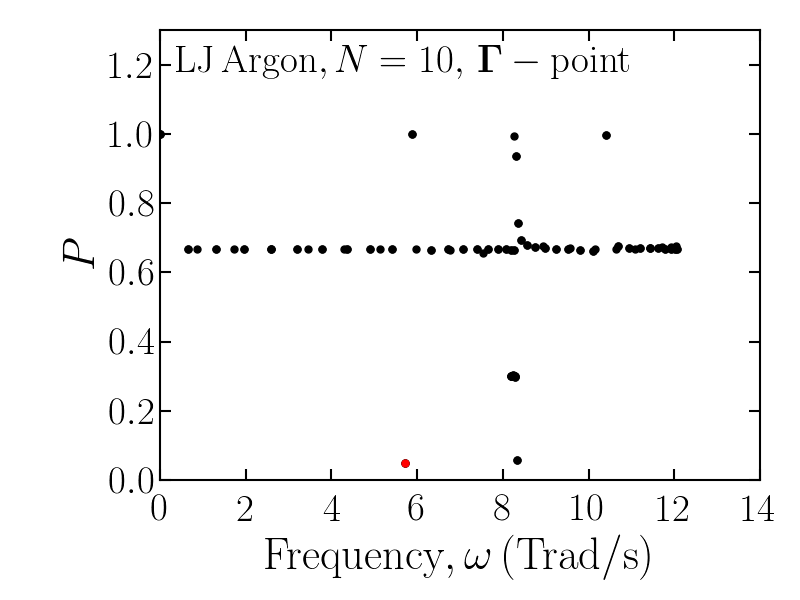}
    \end{subfigure}
   	\\
    
 	\begin{subfigure}[t]{.03\textwidth}
 	(b)
    \end{subfigure}
  	\begin{subfigure}[t]{.45\textwidth}
    \includegraphics[width=\textwidth,valign=t]{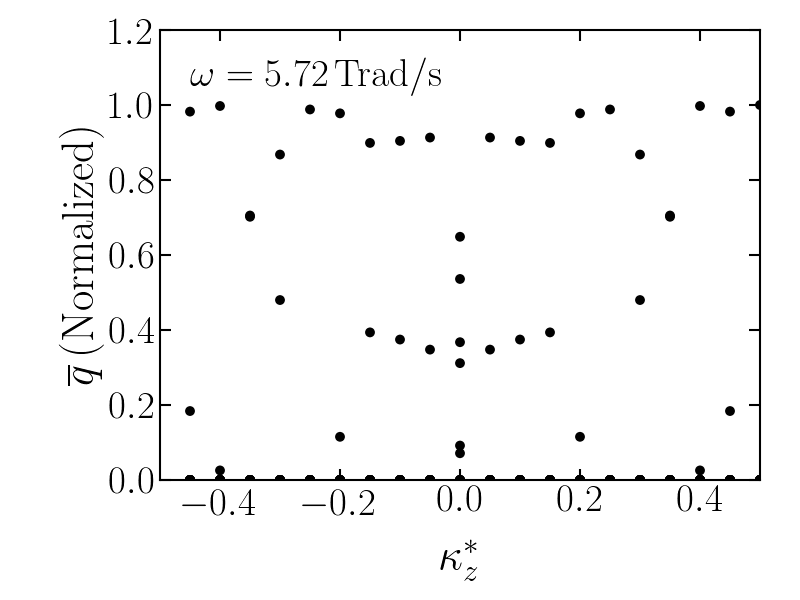}
    \end{subfigure}
    \end{tabular}
    
    \caption{Participation ratios for all $\pmb{\Gamma}$-point modes for the 10 unit cell LJ argon film. Surface modes were identified with participation ratio lower than 0.1 after examining their polarization vectors. The surface mode with a frequency of 5.72 Trad/s is colored red. (b) $\overline{q}$ sweep for the surface mode with a frequency of 5.72 Trad/s. There is no dominant peak, resulting in difficulties for the mapping algorithm.}
    \label{pratio}
\end{figure}

The success of the mapping algorithm is also compromised by the existence of surface phonon modes, which can be identified by their low participation ratios, $P$. This quantity, which has a maximum value of unity, is a measure of the fraction of atoms that participate in a given mode and is defined as \cite{feldman_thermal_1993}
\begin{equation}\label{eq-pratio}
    P(\pmb{\kappa}_\mathrm{2D},\nu_\mathrm{2D}) = \left(\sum_{j,k}\left[\sum_\alpha e_{\mathrm{2D},\alpha,jk}(\pmb{\kappa}_\mathrm{2D},\nu_\mathrm{2D})e_{\mathrm{2D},\alpha,jk}^\dagger(\pmb{\kappa}_\mathrm{2D},\nu_\mathrm{2D})\right]^2\right)^{-1}.
\end{equation}
The participation ratios for the 10 unit cell LJ argon film at the $\pmb{\Gamma}$-point are plotted in Fig. \ref{pratio}(a), where four modes with $P <0.1$ are observed (the two plotted points each correspond to a degenerate mode). 
Upon a closer examination of their polarization vectors, the atomic motions in these modes are mainly at the film surfaces, with little to no motion deep within the film.
Due to the spatial localization of surface modes, the mapping algorithm cannot map them to any 3D mode. 
The application of the mapping algorithm to the surface mode with a frequency of 5.72 Trad/s from Fig.~\ref{pratio}(a) (colored red) results in the $\overline{q}$ sweep plotted in Fig.~\ref{pratio}(b), which shows no distinct peaks.
 
The presence of the above effects at low thicknesses indicates an inherent difference between the 2D phonon modes from their 3D counterparts in LJ argon, going further than a simple limitation on the possible cross-plane wave vectors.
In such cases, the rigorous 2D calculation is necessary to obtain accurate phonon properties.

There is also a source of numerical error that is separate from the anisotropy and the surface modes.
For mapping destinations with the smallest cross-plane wave vector component for a given thickness [i.e., $n= \pm 1$ in Eq.~\eqref{eq-destination}], more mis-mapped modes are present compared to other values of $n$.
This behavior is evident in Figs. \ref{argon-disp}(b) and \ref{argon-disp}(c), which show the mapped 2D dispersions for a ten unit cell film at $\kappa_z^*=0.05$ ($n=1$) and at $\kappa_z^*=0.30$ ($n=6$).
The $\kappa_z^*=0.30$ dispersion has a larger number of 2D mapped modes tracing the 3D branches, but the $\kappa_z^*=0.05$ dispersion has some 3D branches with almost no 2D mapped modes. 
These missing modes are a result of the difficulty of fitting the largest possible wavelength into the thickness of the 2D system.
There are thus mis-mapped modes at other $\kappa_z^*$ because the modes that should have mapped to $\kappa_z^*=0.05$ must end up somewhere.
This effect is present for all film thicknesses at the smallest allowed wave vector, but the increasing number of mapping destinations as the film gets thicker means that the overall effect for the entire 2D dispersion is reduced.

\subsubsection{\label{sec-graph} Graphene}

\begin{figure}
	\begin{tabular}{c c}
	\begin{subfigure}[t]{.03\textwidth}
    (a)
    \end{subfigure}
    
  	\begin{subfigure}[t]{.45\textwidth}
	\includegraphics[width=\textwidth,valign=t]{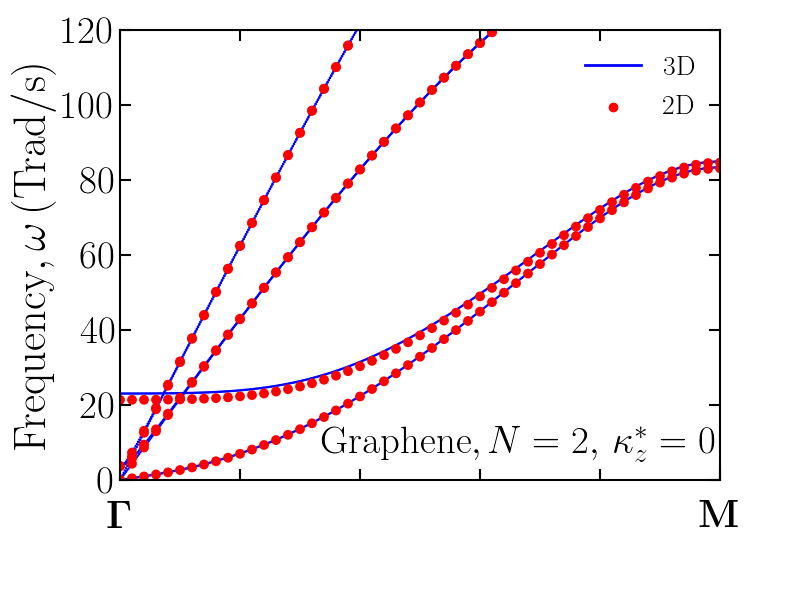}
    \end{subfigure}
   	\\
 	\begin{subfigure}[t]{.03\textwidth}
    (b)
    \end{subfigure}
  	\begin{subfigure}[t]{.45\textwidth}
    \includegraphics[width=\textwidth,valign=t]{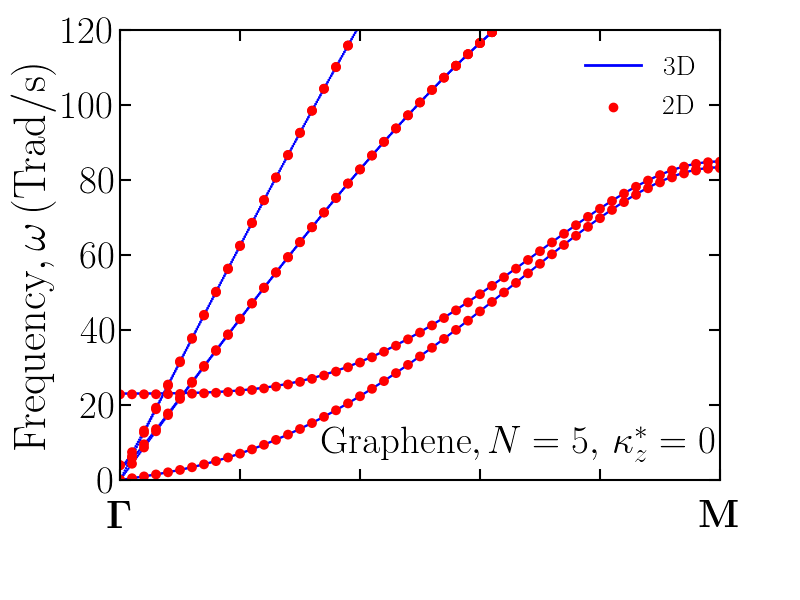}
    \end{subfigure}
    \end{tabular}
    \caption{Graphene dispersions along $\pmb{\Gamma}$-$\mathrm{\mathbf{M}}$ for (a) two unit cell and (b) five unit cell films with $\kappa_z^*=0$. Only the low-frequency modes are shown. Both the two and five unit cell mapped dispersions are in good agreement with the 3D dispersion.}
    \label{graphene-disp}
\end{figure}

The isotropic to anisotropic transition in the LJ argon films can be contrasted with multi-layer graphene and graphite, which are both inherently anisotropic.
Mapped 2D dispersions for the two and five unit cell graphene films are plotted in Figs. \ref{graphene-disp}(a) and \ref{graphene-disp}(b). 
Only the low frequency branches are shown, as higher frequency branches do not vary greatly with respect to $\kappa_z^*$ and are difficult to differentiate.
Also plotted is the bulk graphite $\pmb{\Gamma}-${\bf M} dispersion at $\kappa_z^*=0$. 
The mapped 2D modes in both the two and five unit cell films closely trace the 3D branches. 
The transverse branches are split in both 2D and 3D due to the anisotropy.
A small downward shift in frequency is only observed in the two unit cell film, consistent with the DOS [Fig.~\ref{fig-dos}(b)].
The 2D branches tend to completely map the 3D branches, rather than in the piece-wise manner seen for LJ argon.
This result indicates that the shapes of the 2D branches are very similar to those of the 3D branches, which can be attributed to the weak interactions between the graphene layers.

\section{\label{sec-summary}Summary}
We applied lattice dynamics calculations to study the relationship between phonon modes in corresponding 2D and 3D systems.
Specifically, we developed a mapping algorithm based on normal mode decomposition to link 2D film and 3D bulk dispersions (Sec. \ref{sec-mapping}). 
LJ argon and graphene systems were analyzed due to their respective isotropy and anisotropy in bulk. 

As shown in Figs.~\ref{fig-dos}(a) and \ref{fig-dos}(c), the DOS of LJ argon films converges to the bulk DOS as the thickness increases.
At low thicknesses, however, the 2D and 3D DOS diverge, showing the effect of anisotropy in the films that is not present in bulk.
In contrast, the graphene DOS is very similar to the bulk DOS even at small thicknesses, as shown in Fig. \ref{fig-dos}(b), which is a result of the inherent anisotropy in the bulk material due to weak inter-layer interactions.

Upon application of the mapping algorithm, differences between the 2D and 3D phonon modes were observed in LJ argon. 
At larger thicknesses, the differences are small and the 2D modes correspond strongly to 3D modes with specific cross-plane wave-vector components. Some differences, however, exist in the low-frequency acoustic branches [Figs.~\ref{argon-disp}(a) and \ref{argon-disp}(b)].
At smaller thicknesses, there are significant differences such as downward frequency shifts and branch splitting (Fig. \ref{argon-lj1}).
These effects can be attributed to the emergence of anisotropy and surface modes in the LJ argon films.
The graphene systems, with their weak inter-layer interactions, do not show the same deviations.
Instead, their 2D phonon modes are strongly matched to the 3D phonon modes at specific cross-plane wave-vectors, with only small downward frequency shifts at low frequencies [Figs.~\ref{graphene-disp}(a) and \ref{graphene-disp}(b)].

The observed differences between mapped 2D modes and their 3D counterparts are relevant to predictions of film thermal conductivity. 
The majority of previous calculations have assumed the existence of 3D modes whose mean free paths are modified with a boundary scattering model \cite{jain_phonon_2013,cuffe_reconstructing_2015,fu2017electron,wang_computational_2015}.
In some studies, the number of cross-plane wave vectors was limited by the film thickness \cite{turney_-plane_2010, wang_computational_2015}.
While a treatment based on 3D modes may be appropriate for thicker films or films with weak cross-plane interactions, as we have shown here, they will break down in some very thin films due to the emergence of anisotropy and surface modes. In such cases, an accurate thermal conductivity prediction requires use of the 2D slab unit cell.

\section*{\label{sec:level1}Acknowledgements}

This work was supported by NSF Award DMR-1507325.

The following article has been submitted to the Journal of Applied Physics. After it is published, it will be found at Link.
\bibliographystyle{apsrev4-1}
\bibliography{mappingpaper.bib}

\end{document}